\documentclass[%
 twocolumn,
 superscriptaddress,
 amsmath,amssymb,
 aps,prl,
citeautoscript,
]{revtex4-1}

\usepackage{graphicx}% Include figure files
\usepackage{dcolumn}% Align table columns on decimal point
\usepackage{bm}% bold math

\usepackage[colorlinks = true,
            linkcolor = blue,
            urlcolor  = blue,
            citecolor = blue,
            anchorcolor = blue]{hyperref}
            
\usepackage{fdsymbol}
\usepackage{siunitx}

\begin{document}

\title{Shape and Interaction Decoupling for Colloidal Pre-Assembly}

\author{L. \surname{Baldauf}}
\affiliation{Institute of Physics, University of Amsterdam, 1098XH Amsterdam, The Netherlands.}
\affiliation{Current Address: Department of Bionanoscience, Delft University of Technology, 2629 HZ Delft, The Netherlands.}
\thanks{L.B. and E.G.T. contributed equally.}
\author{E.G.\surname{Teich}}
\affiliation{Applied Physics Program, University of Michigan, Ann Arbor, MI 48109,USA.}
\affiliation{Current Address: Department of Bioengineering, University of Pennsylvania, Philadelphia, PA 19104, USA.}
\thanks{L.B. and E.G.T. contributed equally.}
\author{P. \surname{Schall}}
\affiliation{Institute of Physics, University of Amsterdam, 1098XH Amsterdam, The Netherlands.}
\author{G. \surname{van Anders}}\email{gva@queensu.ca}
\affiliation{Physics, University of Michigan, Ann Arbor, MI 48109, USA.}
\affiliation{Physics, Engineering Physics, and Astronomy, Queen’s University, Kingston ON K7L3N6, Canada.}
\author{L. \surname{Rossi}}\email{l.rossi@tudelft.nl}
\affiliation{Department of Chemical Engineering, Delft University of Technology, 2629 HZ Delft, The Netherlands.}

\date{\today}

\begin{abstract}

Creating materials with structure that is independently controllable
	at a range of scales requires breaking naturally occurring hierarchies.
	Breaking these hierarchies can be achieved via the decoupling of building block attributes from structure during assembly. 
	Here, we demonstrate both geometric and interaction decoupling in pre-assembled colloidal structures of cube-like particles with rounded edges.
	Through computer simulations and experiments, we show that compressing a small number of such cubes in spherical confinement results in clusters with highly reproducible structures that can be  used as mesoscale building blocks to form the next level of structural hierarchy. 
	These clusters demonstrate geometric decoupling between particle shape and cluster structure; namely, for clusters of up to nine particles, the colloidal superballs pack consistently like spheres, despite the presence of shape anisotropy and facets in the cubic-like particles. 
	We confirm that cluster structure is also decoupled from inter-particle interaction, showing that the same structures arise from the spherical confinement of both non-magnetic and magnetic colloidal cubes with strong dipolar interactions.
	To highlight the potential of these superball clusters for hierarchical assembly, we demonstrate, using computer simulations, that clusters of six to nine particles can self-assemble into high-order structures that differ from those of similarly shaped particles without pre-assembly. 
	These results demonstrate  decoupling for anisotropic building blocks that can be further exploited for hierarchical materials development.

\end{abstract} 

\maketitle

\begin{figure}
	\centering
	\includegraphics[width=0.5\textwidth]{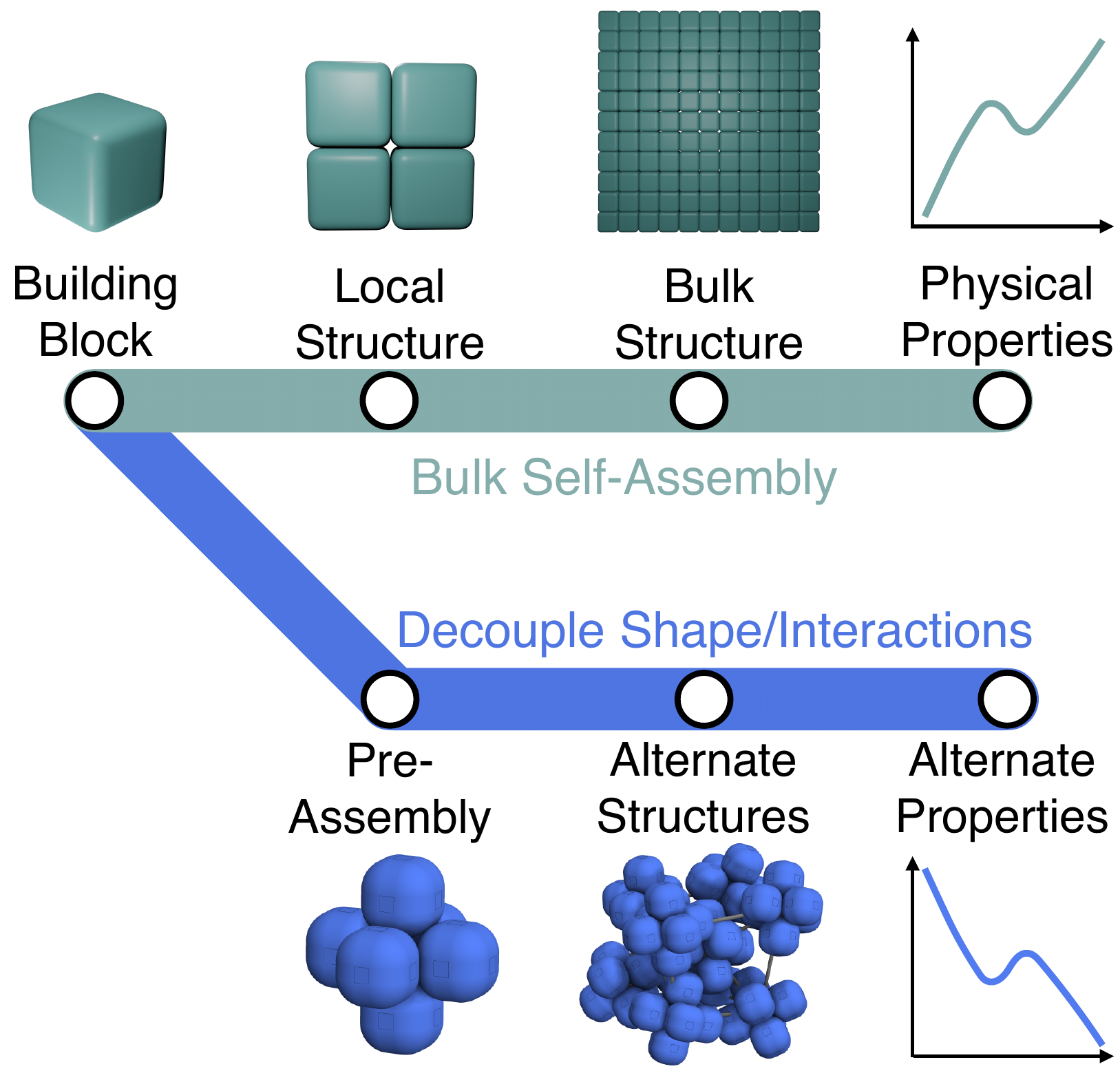}	
	\caption{Schematic illustration of different pathways to hierarchical assembly.  Without intervention, structure and material properties follow directly from building block attributes
			(green ``subway line'' above). Alternate material properties
			require an alternate route, such as pre-assembly (blue subway line) to
			create mesoscale structures that are not found in bulk.}
\label{Figure0}
\end{figure}

A prevailing challenge in materials engineering is to produce materials with controlled, hierarchical structure at several length scales \cite{Lakes1993g,Fratzl2007}.
Hierarchically structured materials can arise in one of two ways.
The first is serendipitous and occurs if target structural motifs at multiple scales arise spontaneously from building blocks \cite{glotzsolomon} with finely-tuned attributes.
Although there have been advances in inverse-design techniques for engineering building block attributes for target structures,\cite{digitalalchemy,miskinjaeger,engent} these techniques have not yet been extended to hierarchical structures.
The other way to achieve hierarchical structure is by intervening during the assembly process to decouple structural outcomes from building block attributes.
Pre-assembling clusters of building blocks into non-bulk motifs provides an avenue to do this.
For spherical building blocks, pre-assembly has been reported in Refs. \citenum{Ducrot:2017cs,Donaldson:2021jz}, in which DNA-mediated and magnetic dipolar interactions were used, respectively, to program hierarchical assembly.
However, generic pre-assembly must also work for shape-anisotropic building blocks, and for these building blocks the key challenge is to decouple their geometry from the resultant structure.
Achieving this decoupling is a pre-requisite for pre-assembly as a viable route for hierarchical materials (see Figure \ref{Figure0} for a schematic illustration).

Here, we demonstrate experimentally that we can induce colloidal shape-anisotropic building blocks to arrange in geometric structures that are not found in bulk self-assembly.
The non-bulk geometric structures we create in experiment can be regarded as a form of ``pre-assembly'' that produces mesoscale building blocks with geometric arrangements that are decoupled from the attributes of the underlying microscopic building blocks.
Using computer simulation, we show that these pre-assembled building blocks self-assemble to produce identical macroscopic structures that differ in their intermediate, mesoscale structure in a way that is dictated by the form of the pre-assembled building block.

We achieve shape decoupling by exploiting novel geometric effects that occur when particles are packed in confinement.
Although the bulk arrangement of shape-anisotropic colloids involves a subtle interplay of multiple forces \cite{Rossi:2015jf,entint}, theoretical predictions \cite{Teich:2016ij} indicate that under idealized conditions, perfectly hard particles in perfectly hard confinement exhibit a range of structural organization that can be at odds with the particle symmetry, and with bulk packing or assembly behavior.
If this putative effect could be realized in experiment through some kind of confinement mechanism, it would be possible to produce mesoscale building blocks within confinement whose structure breaks the emergent hierarchy that exists for free particles in bulk.
Similar mechanisms are known to produce structural arrangement in biology; the packing and crowding of macromolecules in cells \cite{Ellis:2001bu}, the growth of cellular, bacterial and viral aggregates \cite{Hayashi:2004ita,Astrom:2006kq,Gerba:2017bc}, and blood clotting \cite{Cines:2014fka} are a few salient examples.
	
We demonstrate an analogous effect in synthetic colloids using a combination of shape-controlled synthesis and emulsification to confine small numbers of colloids into clusters that defy the arrangement tendencies of their shape and interactions in bulk \cite{Jiao:2009jw,Jiao:2011jh,Ni:2012bc,Meijer:2017hr}.
We show, using experiments and computer simulations, that compressing colloidal superballs with different degrees of sphericity in spherical confinement generates reproducible clusters whose structures are remarkably different than those found in the bulk.   
We confirm experimentally that these clusters are also formed by magnetic colloidal superballs with strong dipolar inter-particle interactions, demonstrating that particle interaction and particle shape can be decoupled via spherical confinement.
We also show via computer simulations that clusters of six to nine particles form higher order assemblies whose structures differ from those observed in bulk assembly of similar particles.
This work introduces a general principle that can be applied to other shape-anisotropic particles to further explore the development of novel materials via hierarchical assembly. 

% ==== Figure 1 ====

\section*{Results and Discussion}

\begin{figure}
	\centering
	\includegraphics[width=0.5\textwidth]{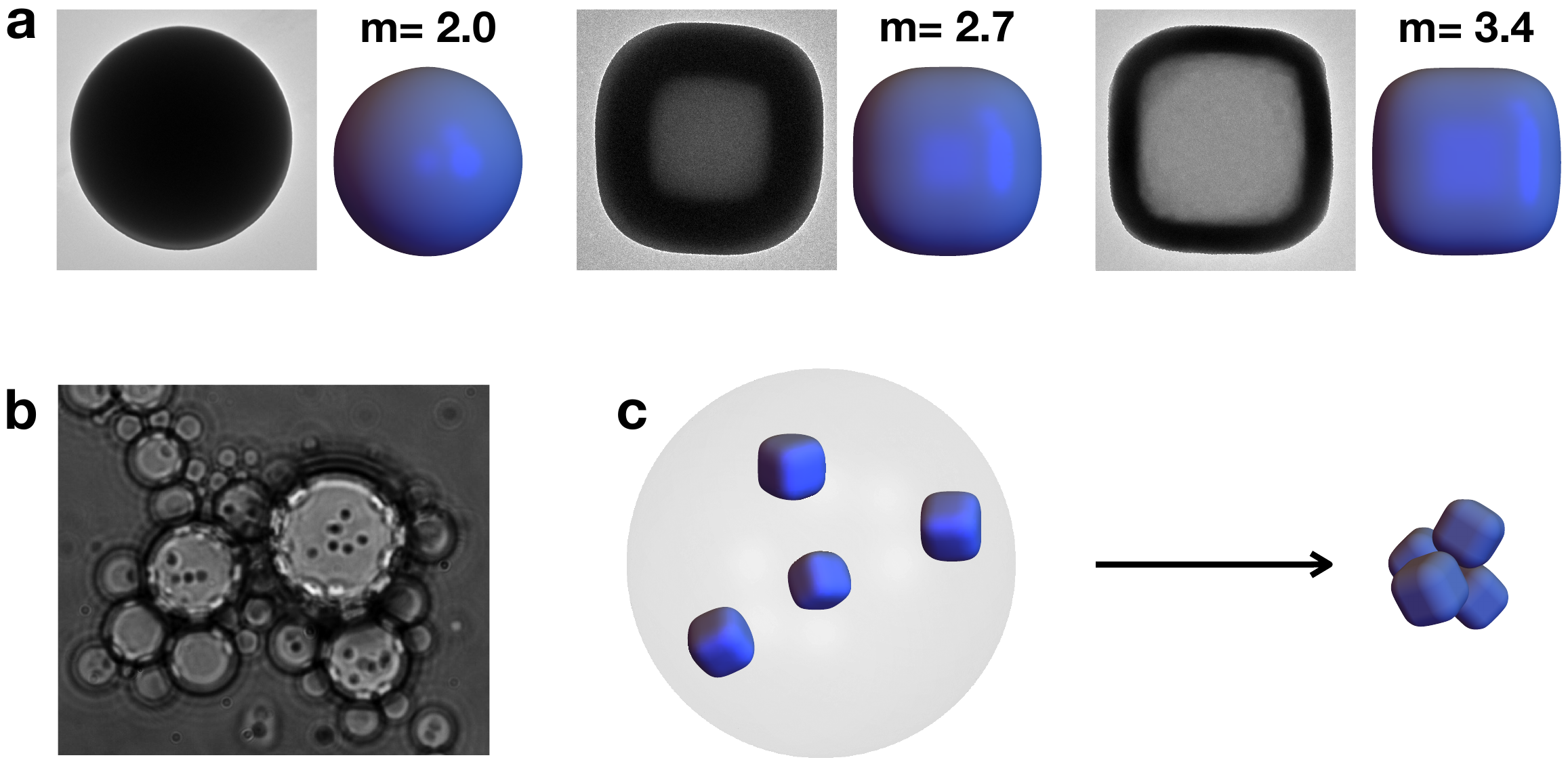}	
	\caption{(a) Transmission electron microscopy images and models of superballs with shape parameter 2.0 (left), 2.7 (middle) and 3.4 (right).(b) Light microscope image of the water-in-oil emulsion containing freely diffusing (spherical) particles.(c) Schematic illustration of the emulsion drying process that compresses the particles together to form a colloidal cluster.
	}
\label{prep}
\end{figure}

%=======================
% Experimental system
%=======================
\noindent \textbf{Experimental system}
\\
\\
Pre-assembly pathways were investigated using colloidal silica cubes with rounded edges. These particles can be modelled as ``superballs'', described by $(x/a)^m + (y/a)^m + (z/a)^m = 1$, where $L=2a$ is the particle's edge length and $m$ is the shape parameter, a value that characterizes the roundness of the particle's corners (see Experimental Methods for details on simulated particles).
This shape is particularly interesting because it smoothly interpolates with $m$ from a sphere at $m=2$ to cubic particles with increasingly sharper corners as $m \rightarrow \infty$.
Here, we used silica particles with three distinct shape parameters: $m=2,  2.7$ and $3.4$ (Figure~\ref{prep}a).
Experimentally, silica superballs were prepared by growing layers of amorphous silica on the surface of cubic hematite particles, with $m$ decreasing with the thickness of the silica layer, as reported in Ref.\citenum{Rossi:2015jf}.
Silica spheres, $m=2$, of size 648~nm were prepared following the classical St\"{o}ber method \cite{Stober:1968tt}, while silica spheres of size 1.2~\si{\micro\meter} were purchased from Bangs Laboratories.
We pre-assembled these particles into small clusters using water-in-oil emulsions following the procedure reported by Cho \textit{et al.} \cite{Cho:2005cp}. 
This method allows for particles to freely diffuse inside the droplets (see Figure~\ref{prep}b) during compression rather than adsorb at the interface affecting their orientations\cite{Soligno:2018dt,Soligno:2016hs} and possibly influencing their final packing geometry. 
In a typical experiment, an aqueous silica dispersion was emulsified in hexadecane using Hypermer as a surfactant (see the Experimental Section for details).
The clusters were formed by compression of slowly evaporating water droplets as shown schematically in Figure~\ref{prep}c.
\\
\\
% ==== Figure 2 ====

%===============
% Comparison to sphere clusters
%===============
\noindent \textbf{Comparison to sphere clusters}
\\
\\
To test the validity of the experimental clustering procedure, we first reproduced clusters of spherical particles using silica spheres.
The results are shown in ESI Figure~S1.
We find that the procedure is reproducible and that the clusters obtained are representative of the different geometries found by Cho \textit{et al.}\cite{Cho:2005cp} for water-in-oil emulsions of silica spheres.
 
\begin{figure}
	\centering
	\includegraphics[width=0.5\textwidth]{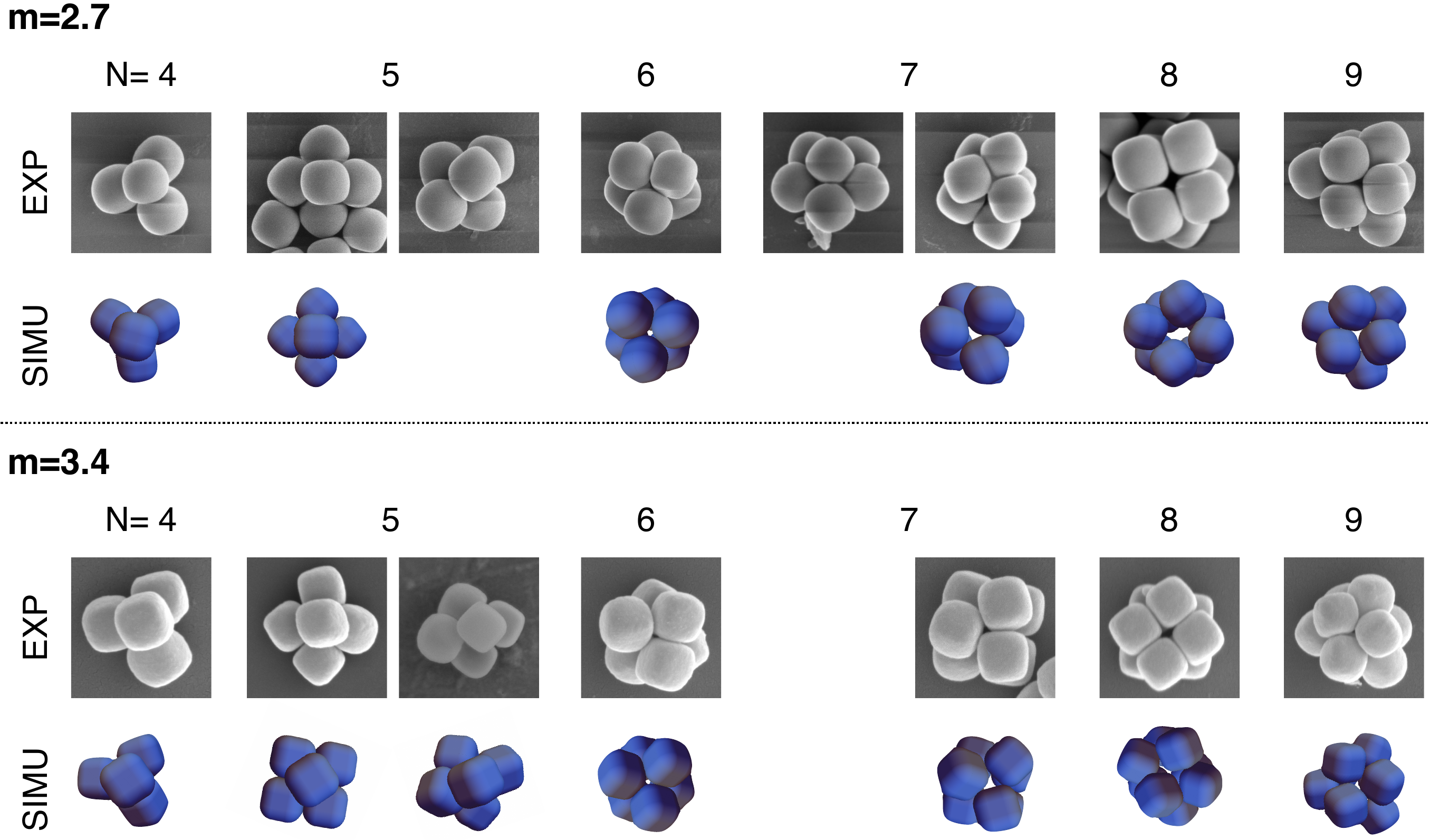}	
	\caption{SEM images of clusters generated from superballs with shape parameters m=2.7 and m=3.4, compared with simulated clusters of superballs with the same shape parameters.
	}
\label{clusters}
\end{figure}

For shape parameters $m=2.7$ and $3.4$, the resulting packing configurations are shown in Figure~\ref{clusters}.
Despite the superballs' lower rotational symmetry, we find that, as with spheres, their packing is uniquely defined, and differs from that of bulk assembly of particles with the same shape parameter.
In bulk, the assembly of superballs can take the form of a $C_{0}$ or $C_{1}$ lattice depending on the $m$ value of the particles \cite{Jiao:2009jw,Jiao:2011jh,Ni:2012bc}. 
%
%This result is supported by simulations that model the packing of similarly shaped cubes in spherical confinement.
%
We find that we consistently obtain clusters whose geometries match those of the corresponding sphere clusters shown in ESI Figure~S1, and found by other experimental \cite{Cho:2005cp, Manoharan:2003hb, Manoharan:2004vs} and simulation works \cite{Teich:2016ij, Lauga:2004im}.
This is surprising because for $m=2.7$ and $3.4$, particles show clear facets (see Figure~\ref{prep}a), and therefore might be expected to pack less like spheres and more like cubes. 
These expectations were also supported by a recent work by some of the authors showing differences in geometry for clusters formed by the Platonic solids. These clusters were generated via computer simulations through compression of a spherical container\cite{Teich:2016ij}, a method which can be directly related to the experimental evaporation of water-in-oil emulsion droplets that we are employing. 
In their work, Teich \textit{et al.} showed that clusters of perfect cubes ($m=\infty$) present a common structure with clusters of spheres only for clusters of $N=4$ and $5$ particles. 
They found that clusters of higher numbers of perfect cubes are not similar to clusters of spheres according to a similarity metric comparing vectors of descriptive Steinhardt order parameters. Instead, these clusters tend to organize into layers  with primarily orthorhombic, monoclinic, or cubic symmetries. 
Thus, these cluster structures are highly influenced by the cubic shape of their constituent particles.
This coupling between shape and structure is in contrast to the decoupling found in this work for cubes with rounded edges (shape parameter $m=2.7$ and $3.4$).

% ==== Figure 3 ====
%\begin{figure*} for two column figure!!!
For clusters of $N=5$ particles ($m=2.7$ and $3.4$) we find degenerate structures: square pyramid and triangular dipyramid. 	%
Although experimental works on sphere clusters of both water-in-oil (w-i-o)\cite{Cho:2005cp} and oil-in-water (o-i-w)\cite{Manoharan:2003hb} emulsions have reported only triangular dipyramid clusters, we consistently find both geometries in our experiments with spherical particles (see ESI Figure~S1). 
\\
\\
%===============
% Comparison to simulation
%===============
\noindent \textbf{Comparison to simulation}
\\
\\
To check the robustness of these experimental results, we computationally generated dense clusters of hard superballs with the experimental shape parameters $m=2.7$ and $3.4$ using isobaric Monte Carlo simulations identical in protocol to those described in Ref. \citenum{Teich:2016ij}.
	We encased our particles inside a spherical container, and enforced confinement by rejecting trial particle moves if they resulted in any overlaps between particles and the surrounding spherical wall. We also rejected trial particle moves if they resulted in any overlaps between particles. We induced compression of the spherical container by exponentially increasing system pressure to a putatively high value during the simulation. We ran simulations of $N=4-9$ particles for each shape parameter, and for every $(N,m)$ statepoint we ran 50 compression simulations. We chose the densest cluster achieved at each $(N,m)$ statepoint for further analysis and comparison to experimental results (see Experimental Section).
	
A comparison between representative SEM images of clusters of superballs with $m$ values of $2.7$ and $3.4$ and the corresponding clusters achieved via simulation is included in Figure~\ref{clusters}.
	We find good qualitative agreement between experiments and simulations.
	For each ($N,m$) statepoint, a majority (over 80\%) of our 50 replicate simulations resulted in approximately the same final cluster structure, and this structure (the densest of which is shown in Fig. \ref{clusters}) approximately matches experimental results. 
	A notable exception is the case of $N=5$. 
	For ($N,m$) = (5,2.7), we find only the square pyramid in our simulations (with all four base particles aligned face-to-face in 44 replicates, or with one base particle misaligned in the remaining replicates). 
	For ($N,m$) = (5,3.4), however, we find a skewed square pyramid in only 24/50 replicate simulations, and an approximate triangular bipyramid in the remaining 26/50 simulations. 
	The triangular bipyramid is the denser final structure, since the density of the densest triangular bipyramid is larger than the density of the densest square pyramid by $\approx 3.7 \times 10^{-4}$.
	%
	% We speculate that, even if low (about 4~\%), polydipersity in the shape of the particles gives rise to these two degenerate structures in our experiments.
	%
	
\begin{figure}
	\centering
	\includegraphics[width=0.5\textwidth]{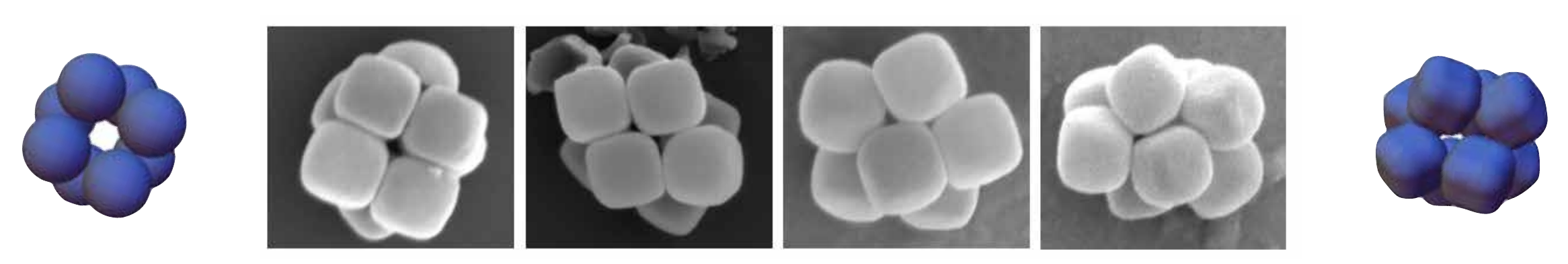}	
	\caption{SEM images of isomeric clusters of 8 particles with shape parameter $m=3.4$. The model on the left shows a twisted-square structure as obtained with spheres and the model on the right shows a compressed structure similar to a snub disphenoid as obtained with superballs with $m =$ 3.4. 
	}
\label{transition}
\end{figure}

\begin{figure}
	\centering
	\includegraphics[width=0.5\textwidth]{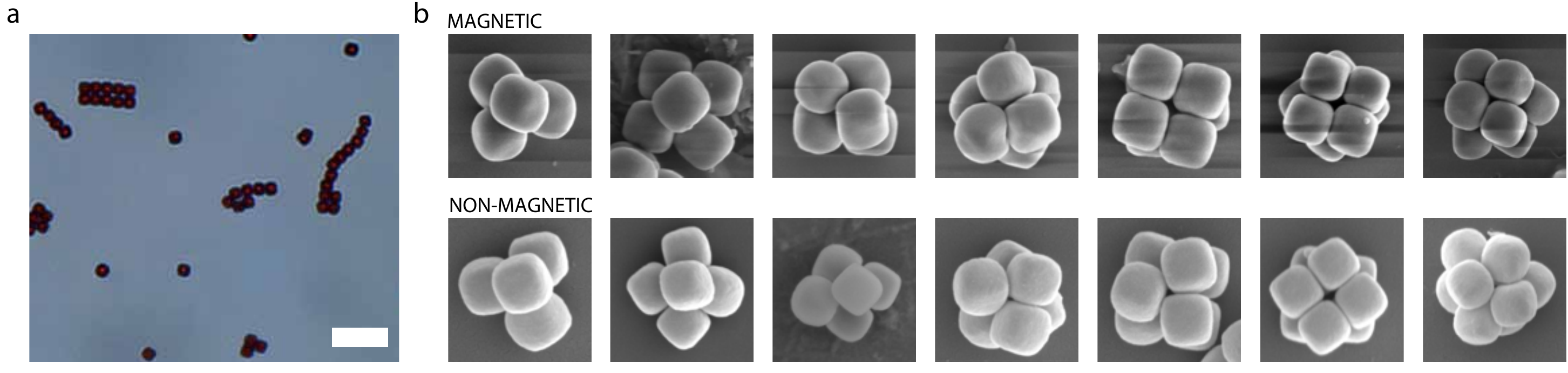}	
	\caption{(a) Optical microscope image of the magnetic cubes forming 2D dipolar structures in a zero-field environment. Scale bar is 8~\si{\micro\meter}. (b) Representative SEM images of clusters prepared using magnetic (top row) and non-magnetic (bottom row) superball particles. For all clusters we see consistent geometries which depend only on the number of constituent particles.}
\label{magnetic}
\end{figure}

	For $N=8$, the prevalent structure found experimentally for superballs with shape parameters $m=2.7$ and $3.4$ is the twisted-square (Figure~\ref{clusters}) also found in spheres in the experiments by Cho \textit{et al.} \cite{Cho:2005cp} and simulations by Teich \textit{et al.}\cite{Teich:2016ij}. 
	In the computer simulations of particles with the same shape parameters (also in Figure~\ref{clusters}), we obtain clusters with a similar but compressed structure. 
	The same compressed twisted-square structures were also present in experiments and, although less abundant than the twisted-square clusters, they were found in all samples.
	Figure~\ref{transition} shows the compressed structures as a transition between the twisted-square structure and a snub disphenoid, a structure which minimizes the second moment of the mass distribution.
	This structure is largely found in clusters of spheres obtained from o-i-w emulsion droplets of uncharged particles\cite{Manoharan:2003hb}.
	It seems that, although less favorable than the twisted-square, the intermediate structures that were also predicted by our computer simulations are accessible by our systems.
	Computer simulations of particles with higher shape parameters (detailed in the next section and Supporting Figure S2) show that for clusters with $N=8$ particles there is a transition from twisted-square for $m=2$ to snub disphenoid for $m=2.7$ and $3.4$ and back to twisted-square for $m=6$ and $8$. This transition is solely dependent on the shape parameter.
	Since the transition is shape dependent, we can hypothesize that the presence of isomers in the experiments is attributable to particle polydispersity in shape.
% ==== Figure 4====
%\begin{figure*} for two column figure!!!
\\
\\
 %===============
% Simulation of higher $m$ values
%===============
\noindent \textbf{Simulation of higher $m$ values}
\\
\\
To further explore the importance of particle shape in determining cluster geometry, we performed computer simulations of particles with higher $m$ values, specifically $m=6$ and $m=8$.
	Particles with such high shape parameters are not accessible using hematite particles and are in general more difficult to obtain experimentally in the micron-size range.
	The results are reported in Supporting Figure~S2.
	We find that, despite sharper particle edges, clusters of superballs with $m=6$ and $m=8$ show densest structures that are equivalent to those already found for lower $m$ values. 
	As mentioned above, both $m=6$ and $m=8$, clusters with $N=8$ particles show the twisted-square geometry most similar to that of clusters of spheres, although for $m=6$ and $m=8$ the two 4-particle layers forming the clusters are off-centered by angles less than 45 degrees. 
\\		
\\
 %===============
% Clusters of magnetic superballs
%===============
\noindent \textbf{Clusters of magnetic superballs}
\\
\\
While packing considerations are clearly important to determine the final resulting cluster geometry, the question remains whether other factors, such as inter-particle interactions, play a role in influencing the final structure.
	Recent computer simulations suggest that for magnetic particles with $m=4$ the packing is still dominated by geometry rather than dipolar interaction\cite{Donaldson:2021jz}.
	To test this result experimentally, we prepared clusters of magnetic particles with strong dipolar inter-particle interactions. 
	We used core-shell hematite-silica particles with comparable shape parameters prepared as described in Ref.~\citenum{Rossi:2018ef}. 
	The procedure to prepare the magnetic core-shell particles is identical to that used for the preparation of the non-magnetic particles, with only one difference: the internal magnetic hematite core is not dissolved once the silica shell has been deposited on its surface.
	The resulting particles have a superball shape, an external silica surface identical to that of the silica superballs, and a permanent magnetic dipole moment strong enough ($\mu_{p}\approx~3\times10^{-15}$~\si{\ampere\meter\squared}) to induce dipolar structure formation even in the absence of an external magnetic field\cite{Rossi:2018ef}.
	An example of such dipolar structures can be seen in the optical microscope image in Figure~\ref{magnetic}a.
	%

% ==== Figure 5====
%\begin{figure*} for two column figure!!!

	We find that clusters made using magnetic superballs again have the same geometry as those formed by (non-magnetic) silica superballs, confirming that shape is the crucial parameter in determining the final cluster geometry, and that the experimental spherical confinement procedure is capable of decoupling inter-particle interaction from resultant cluster structure. 
\\
\\	
\\
\\	
%===============
% Hierarchical self-assembly
%===============
\noindent \textbf{Hierarchical self-assembly}
\\
\\
To test whether the clusters produced in experiment provide viable candidates for hierarchical self-assembly, we modeled the experimentally produced clusters as rigid collections of hard colloids in Monte Carlo simulations.
We performed simulations of $N_c=343$ clusters and observed the self-assembly of ordered structures over a range of fixed densities, using the hard-particle Monte Carlo (HPMC) plugin \cite{Anderson2015a} for HOOMD-Blue \cite{Anderson2008b}. We used a standard $NVT$ Monte Carlo simulation protocol that has been widely adopted in other investigations (e.g.\ \cite{trunctet,zoopaper,entint,engent}).

\begin{figure}
	\centering
	\includegraphics[width=0.5\textwidth]{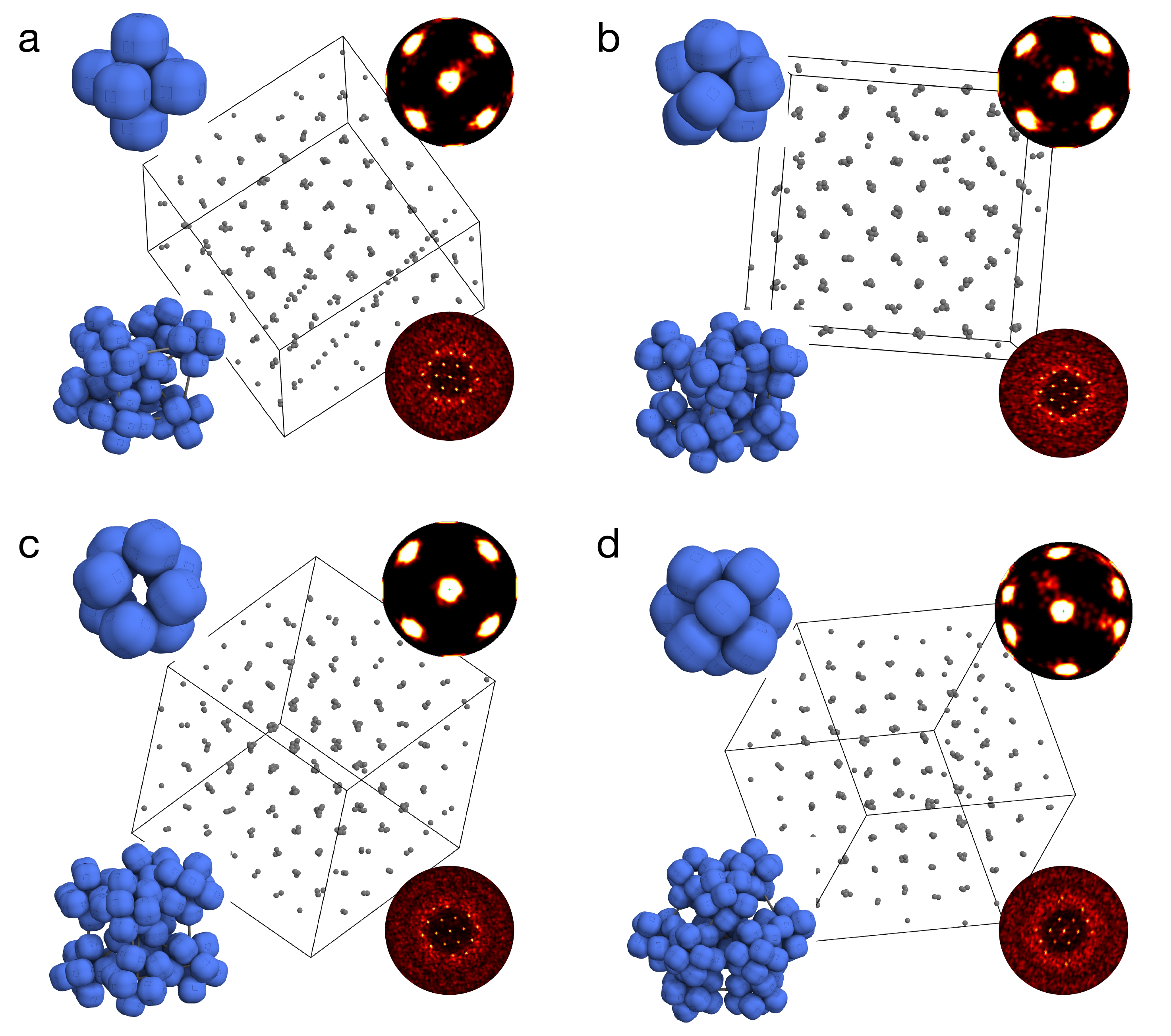}	
	\caption{Superball clusters form hierarchical assemblies.
		Each panel displays an assembly snapshot of $m=2.7$ superball clusters in the center, and an image of the superball cluster in the upper left corner. 
		In all assembly snapshots only the cluster centers are shown as gray spheres, for clarity.
		Bond-orientational order diagrams (BODs, calculated for cluster centers) are shown in the upper right corner of each panel, and indicate the local environment of clusters surrounding each cluster center.
		Diffraction patterns calculated for all superballs are displayed in the lower right corner of each panel, showing ordering at long range (small $\mathbf{k}$) and disorder at short range (large $\mathbf{k}$).
		Diffraction patterns are oriented to show the 3-fold rotational symmetry of each structure.
		Finally, images depicting examples of the local arrangement of clusters in each structure are shown in the lower left corner of each panel.
		The self-assembled systems shown here consist of cluster centers arranged in a BCC structure for (a) $N=6$, $\phi=0.4$, (b) $N=7$, $\phi=0.4$, and (c) $N=8$, $\phi=0.42$; and an HCP structure for (d) $N=9$, $\phi=0.4$.
		For the BCC structures, the local arrangement of clusters displayed is a central cluster surrounded by its 8 closest neighbors, forming an approximate cube.
		For the HCP structure, the local arrangement of clusters displayed is a central cluster surrounded by its 12 nearest neighbors, forming an approximate anticuboctahedron.}
\label{SA}
\end{figure}

In contrast to the structures typically observed in the self-assembly of cubic monomers \cite{Ni:2012bc,Jiao:2009jw,Jiao:2011jh, Meijer:2017hr,Rossi:2015jf},
we found that superball clusters self-assembled into bulk BCC or HCP structures, structures more typical of the self-assembly of monomers of other shapes.
We also observed that pre-assembled clusters self-assembled at lower densities than typically observed for monomer self-assembly. For example, we observed $N=6,8,9$ clusters assembled hierarchically at packing densities as low as 38\%, and $N=7$ at 40\%, whereas monomer self-assembly is more typically observed at densities of 50\% or more\cite{Ni:2012bc}.
Fig. \ref{SA} shows examples of these self-assembled structures for clusters consisting of $N=6,7,8,9$ superballs with $m=2.7$.
The hiearchical nature of the self-assembly is most clearly evident in the simulated diffraction images (lower right inset in each panel in Fig.\ \ref{SA}). The diffraction peaks at low wavenumber show the long range order of the BCC arrangement of cluster centres.
In contrast, the large wavenumber ring reflects the short-range disorder of the cluster orientations.
Crucially, we consistently observed this form of bulk self-assembly for clusters containing different numbers of particles.
Because of the geometric arrangement of the particles within clusters, clusters of different sizes have distinct organization. This means that although we observed consistent long-range order in our cluster assemblies, the short-range order is considerably different across different clusters.
We also note that, in addition to the change in structural order we observe for the hierarchical assembly of clusters, the $>10\%$ reduction in required density for the self-assembly of pre-assembled clusters versus monomers suggests that pre-assembly could be a route to lower density structures.
%
% ==== Figure 6====
%\begin{figure*} for two column figure!!!

%=============
%=============
% Discussion
%=============
%=============
\section*{Conclusions}

We demonstrated that the emulsification of shape-anisotropic colloids produces clusters with geometric arrangements, not observed in the bulk self-assembly or packing of similarly shaped particles\cite{Ni:2012bc,Jiao:2009jw,Jiao:2011jh, Meijer:2017hr,Rossi:2015jf}, that are at odds with the underlying particle symmetry.
	We found that treating these clusters as pre-assembled building blocks yields hierarchically structured self-assembly in computer simulations.
	Our experimental findings that similar clustering effects persist for particles, that are shape-anisotropic and have magnetic interactions, demonstrate that our approach can be used to suppress the effects of both shape and interaction.
	Collectively, our results demonstrate that our emulsification method produces a form of shape and interaction decoupling that can be exploited as a route to the general pre-assembly of building blocks for the design of hierarchically structured materials.
	The experiments we performed here involved clusters of up to nine building blocks.
	Simulations predict that similar effects can be achieved in clusters with dozens of building blocks \cite{Teich:2016ij}, and experiments suggest that non-bulk behavior can be produced in clusters containing hundreds of thousands of building blocks\cite{deNijs:2014fb}.
	These results suggest that the pre-assembly approach we demonstrate here can extend to building block geometries, forms of interaction, and cluster sizes well beyond those we studied.

	Moving beyond the present proof-of-principle requires achieving scale and generalization. For scale, we note that prior work shows that emulsification can be followed by the successful separation of clusters, usually via density gradient centrifugation\cite{Manoharan:2003hb}.
	In terms of generalization, on the theory side, building blocks with geometries of the sort presented here can be incorporated into inverse design approaches.
	Pre-assembled building blocks that are sufficiently small to maintain Brownian motion, can be included in an inverse design approach like digital alchemy \cite{digitalalchemy,engent}.
	For larger building blocks that become granular, pre-assembled building blocks formed from densely packed clusters could be used to generate ``mutations'' in evolutionary design approaches \cite{miskinjaeger}.
	On the experimental side, our results showing that similar effects can be produced in particles with shape-anisotropy and a magnetic core suggests that further application of this method could be immediately refined to produce pre-assembled building blocks with magnetic patches\cite{Donaldson:2021jz}.
	Additional experiments with differently shaped particles will yield pre-assembled of colloidal clusters with novel geometric complexity and specific surface functionalization, forming the basis for further development of hierarchically structured materials.
	%

%=========
%=========
% Methods
%=========
%=========
\section*{Methods}

% ==== Preparation of silica superballs ====
\noindent \textbf{Preparation of silica superballs.}
\\
\\
Silica superballs with different shape parameters were prepared employing the procedure described in a previous work by some of the authors \cite{Rossi:2015jf}. Silica spheres (superballs with $m=2$) were prepared following the well known St\"{o}ber method \cite{Stober:1968tt}. 
\\
\\
% ==== Preparation of superball clusters. ====
\noindent \textbf{Preparation of superball clusters.}
\\
\\
Colloidal clusters were prepared from w-i-o emulsions following a modified version of Cho \textit{et al.}'s procedure \cite{Cho:2005cp}. The oil phase was prepared by dissolving 0.0144~g Hypermer 2296 (2266-LQ-(MV), donated by Croda)  in 4.808~g hexadecane (ReagentPlus 99\%, Sigma Aldrich). The water phase consisted of 780~\si{\micro\liter} of particle dispersion (silica or hematite-silica particles) at an approximate concentration of $7.5 \times 10^{-9}$ colloids/mL. The two phases were emulsified using a Silverson L5M-A emulsifier rotating first at 8000~rpm for 40~s and then at 9500~rpm for an additional 40~s. After emulsification the vial containing the emulsion was placed in a heated glycerol bath at a constant temperature of 100~$^{\circ}$C. The emulsion was magnetically or mechanically stirred throughout the whole drying process which typically lasted  one hour. The evaporation process was monitored in time by optical microscopy observations. The newly formed clusters were then washed by sedimenting and redispersing in hexane containing a low concentration ($\leq 0.1~\%wt$) of Hypermer. Clusters can then be stored in hexane or dried and redispersed in a $0.1~\%wt$ SDS aqueous solution. 
\\
\\
% ==== Electron microscopy. ====
\noindent \textbf{Electron microscopy.}
\\
\\
Particle size and shape were determined by transmission electron microscopy (TEM, Philips TECNAI12). Samples were prepared by drying drops of diluted dispersions on carbon and polymer coated copper grids. Cluster geometries were studied by scanning electron microscopy (FEI Helios Nanolab 600). Samples were prepared by drying drops of dispersions on carbon and polymer coated copper grids or small pieces of silicon wafer, which were then mounted on 15~mm SEM stubs using conductive tape. All SEM samples were sputter coated (Leica EM ACE600) with a layer of 5~\si{\nano\meter} Cr and 5~\si{\nano\meter} Au. 
\\
\\
% ==== Optical microscopy. ====
\noindent \textbf{Optical microscopy.}
\\
\\
Emulsions and clusters were imaged using a Zeiss Axio Imager A1 upright microscope with a Plan NEOFLUAR 100$\times$ oil immersion objective. Dispersions were placed in flat rectangular borosilicate glass capillaries (CM Scientific 0.2~mm~$\times$~2~mm~$\times$~4~cm), sealed to a microscope slide with wax. 
\\
\\
% ==== Simulations of colloidal clusters. ====
\noindent \textbf{Simulations of colloidal clusters.}
\\
\\
We computationally generated dense clusters of superballs via isobaric Monte Carlo simulations using spherical confinement, identically to the protocol described in Ref. \citenum{Teich:2016ij}. We used the hard particle Monte Carlo (HPMC) \cite{Anderson2015a} method in the open-source simulation toolkit HOOMD-blue \cite{Anderson2008b, Glaser2015}. The computational workflow and data management were facilitated by the signac data management framework \cite{Adorf2016a, Adorf2017a}. We visualized our results using the software package Injavis \cite{injavis}. For each $(N,m)$ state point, we placed $N$ superballs with shape parameter $m$ inside a sphere. We rejected trial particle translations or rotations if they resulted in any particle overlaps or overlaps between particles and the encasing sphere. We enforced increasing spherical confinement by raising dimensionless pressure exponentially from a minimum value of 0.1 to a maximum value of 500 in $10^4$ steps. Dimensionless pressure is defined as $p^*=\beta p l^3$, where $p$ is pressure and $l=L$ is the superball edge length as defined in the main text. At each pressure value, we allowed the system to equilibrate for $10^3$ HPMC timesteps. Throughout each simulation, we tuned particle translation and rotation move sizes, as well as system volume move sizes, to maintain move acceptance ratios of approximately 0.2.
    We ran simulations of $N=4-9$ particles for each shape parameter, and for every $(N,m)$ statepoint we ran 50 compression simulations. We chose the densest cluster achieved at each $(N,m)$ statepoint for further analysis and comparison to experimental results.
    
    We approximated each superball, parameterized by $(L,m)$, as a spherocube, or a cube with a sphere swept around its edges, parameterized instead by $(\sigma, d)$. $\sigma$ is the side length of the cube and $d$ is the diameter of the sweeping sphere. We used the approximating spherocubes as the particle shapes in our simulations because the spheropolyhedron overlap check was already implemented in HPMC, and because spherocubes are very good approximations of superballs. To generate each approximation, we followed the prescription laid out in Appendix C of Ref. \citenum{Marechal2012}; we refer the interested reader to that paper for more detail. In brief, we found $(\sigma, d)$ such that the corresponding spherocube had the minimal Hausdorff distance from the superball defined by each $(L,m)$ value of interest. The Hausdorff distance $d(A,B)$ between bodies $A$ and $B$ is defined as:
    	\begin{align}
    		d(A,B) &= \text{max} \{ d'(A,B), d'(B,A)\} \notag \\
    		d'(A,B) &\equiv \text{max}_{\bf{x}\in A} \text{min}_{\bf{y} \in B} \vert \bf{x} - \bf{y} \vert \notag
    	\end{align}
	
	\noindent Calculation of the Hausdorff distance between a superball and spherocube, $d(L,m,\sigma,d)$, is straight-forward, as it is only necessary that one finds the maximal distance over three special pairs of $\{ \bf{x},\bf{y} \}$ points. (The three pairs lie along the 2-fold, 3-fold, and 4-fold symmetry axes of the aligned spherocube/superball pair respectively.) We set $L=1$ and found the following minimal Hausdorff distances between each $(L,m)$ superball and its optimally matching $(\sigma, d)$ spherocube:
	
	\begin{align}
		d(1,2.7,0.209,0.795) &\approx 0.0019 \notag \\
		d(1,3.4,0.347,0.658) &\approx 0.0024 \notag \\
		d(1,6.0,0.604,0.401) &\approx 0.0024 \notag \\
		d(1,8.0,0.696,0.308) &\approx 0.0021 \notag
	\end{align} 
	
	\noindent These were the spherocube parameters we used in our simulations. \\
	
\noindent \textbf{Hierarchical Assembly.}
    \noindent We tested for the possibility of hierarchical self-assembly by treating simulated colloidal clusters as hard, pre-assembled building blocks that we simulated using the hard particle Monte Carlo (HPMC) \cite{Anderson2015a} method in the open-source simulation toolkit HOOMD-blue \cite{Anderson2008b, Glaser2015}. We initialized systems of $N=343$ clusters at low density which we thermalized for $10^5$ Monte Carlo (MC) sweeps, and compressed systems to target packing fractions of $38\%$ and above over $5\times 10^6$ MC sweeps. We examined final simulation snapshots for structural order by examining simulated diffraction patterns and bond-order diagrams via the software package Injavis \cite{injavis}. We did not observe conclusive evidence of self-assembly on the timescale of our simulations for clusters of particles with $m=3.4$. For $m=2.7$ we observed hierarchical self-assembly of clusters of $N=6,7,8,9$ particles.
	
%=================
%=================
% Acknowledgments 
%=================
%=================

\acknowledgments
The authors thank Marije Kronemeijer for the preparation of magnetic superball clusters. L.R. acknowledges the Netherlands Organisation for Scientific Research (NWO) for financial support through a Veni grant (680-47-446). L.B. acknowledges the support by a Studienstiftung des Deutschen Volkes research grant. G.v.A. acknowledges the support of the Natural Sciences and Engineering Research Council of Canada (NSERC) grants RGPIN-2019-05655 and DGECR-2019-00469. E.G.T acknowledges support from the National Science Foundation Graduate Research Fellowship Grant DGE 1256260 and a Blue Waters Graduate Fellowship. This research is part of the Blue Waters sustained petascale computing project, which is supported by the National Science Foundation (awards OCI-0725070 and ACI-1238993) and the state of Illinois. Blue Waters is a joint effort of the University of Illinois at Urbana-Champaign and its National Center for Supercomputing Applications. This research also utilized computational resources and services supported by Advanced Research Computing at the University of Michigan, Ann Arbor.

\providecommand{\latin}[1]{#1}
\makeatletter
\providecommand{\doi}
  {\begingroup\let\do\@makeother\dospecials
  \catcode`\{=1 \catcode`\}=2 \doi@aux}
\providecommand{\doi@aux}[1]{\endgroup\texttt{#1}}
\makeatother
\providecommand*\mcitethebibliography{\thebibliography}
\csname @ifundefined\endcsname{endmcitethebibliography}
  {\let\endmcitethebibliography\endthebibliography}{}


\begin{mcitethebibliography}{38}
\providecommand*\natexlab[1]{#1}
\providecommand*\mciteSetBstSublistMode[1]{}
\providecommand*\mciteSetBstMaxWidthForm[2]{}
\providecommand*\mciteBstWouldAddEndPuncttrue
  {\def\EndOfBibitem{\unskip.}}
\providecommand*\mciteBstWouldAddEndPunctfalse
  {\let\EndOfBibitem\relax}
\providecommand*\mciteSetBstMidEndSepPunct[3]{}
\providecommand*\mciteSetBstSublistLabelBeginEnd[3]{}
\providecommand*\EndOfBibitem{}
\mciteSetBstSublistMode{f}
\mciteSetBstMaxWidthForm{subitem}{(\alph{mcitesubitemcount})}
\mciteSetBstSublistLabelBeginEnd
  {\mcitemaxwidthsubitemform\space}
  {\relax}
  {\relax}

\bibitem[Lakes(1993)]{Lakes1993g}
Lakes,~R. Materials with structural hierarchy. \emph{Nature} \textbf{1993},
  \emph{361}, 511--515\relax
\mciteBstWouldAddEndPuncttrue
\mciteSetBstMidEndSepPunct{\mcitedefaultmidpunct}
{\mcitedefaultendpunct}{\mcitedefaultseppunct}\relax
\EndOfBibitem
\bibitem[Fratzl and Weinkamer(2007)Fratzl, and Weinkamer]{Fratzl2007}
Fratzl,~P.; Weinkamer,~R. Nature's hierarchical materials. \emph{Progress in
  Materials Science} \textbf{2007}, \emph{52}, 1263--1334\relax
\mciteBstWouldAddEndPuncttrue
\mciteSetBstMidEndSepPunct{\mcitedefaultmidpunct}
{\mcitedefaultendpunct}{\mcitedefaultseppunct}\relax
\EndOfBibitem
\bibitem[Glotzer and Solomon(2007)Glotzer, and Solomon]{glotzsolomon}
Glotzer,~S.~C.; Solomon,~M.~J. Anisotropy of Building Blocks and Their Assembly
  Into Complex Structures. \emph{Nat. Mater.} \textbf{2007}, \emph{6},
  557--562\relax
\mciteBstWouldAddEndPuncttrue
\mciteSetBstMidEndSepPunct{\mcitedefaultmidpunct}
{\mcitedefaultendpunct}{\mcitedefaultseppunct}\relax
\EndOfBibitem
\bibitem[van Anders \latin{et~al.}(2015)van Anders, Klotsa, Karas, Dodd, and
  Glotzer]{digitalalchemy}
van Anders,~G.; Klotsa,~D.; Karas,~A.~S.; Dodd,~P.~M.; Glotzer,~S.~C. {Digital
  Alchemy for Materials Design: Colloids and Beyond}. \emph{ACS Nano}
  \textbf{2015}, \emph{9}, 9542--9553\relax
\mciteBstWouldAddEndPuncttrue
\mciteSetBstMidEndSepPunct{\mcitedefaultmidpunct}
{\mcitedefaultendpunct}{\mcitedefaultseppunct}\relax
\EndOfBibitem
\bibitem[Miskin and Jaeger(2013)Miskin, and Jaeger]{miskinjaeger}
Miskin,~M.~Z.; Jaeger,~H.~M. Adapting granular materials through artificial
  evolution. \emph{Nat. Mater.} \textbf{2013}, \emph{12}, 326--331\relax
\mciteBstWouldAddEndPuncttrue
\mciteSetBstMidEndSepPunct{\mcitedefaultmidpunct}
{\mcitedefaultendpunct}{\mcitedefaultseppunct}\relax
\EndOfBibitem
\bibitem[Geng \latin{et~al.}(2019)Geng, van Anders, Dodd, Dshemuchadse, and
  Glotzer]{engent}
Geng,~Y.; van Anders,~G.; Dodd,~P.~M.; Dshemuchadse,~J.; Glotzer,~S.~C.
  Engineering Entropy for the Inverse Design of Colloidal Crystals from Hard
  Shapes. \emph{Science Advances} \textbf{2019}, \emph{5}, eaaw0514\relax
\mciteBstWouldAddEndPuncttrue
\mciteSetBstMidEndSepPunct{\mcitedefaultmidpunct}
{\mcitedefaultendpunct}{\mcitedefaultseppunct}\relax
\EndOfBibitem
\bibitem[Ducrot \latin{et~al.}(2017)Ducrot, He, Yi, and Pine]{Ducrot:2017cs}
Ducrot,~{\'E}.; He,~M.; Yi,~G.-R.; Pine,~D.~J. {Colloidal alloys with
  preassembled clusters and spheres}. \emph{Nature Materials} \textbf{2017},
  \emph{16}, 652--657\relax
\mciteBstWouldAddEndPuncttrue
\mciteSetBstMidEndSepPunct{\mcitedefaultmidpunct}
{\mcitedefaultendpunct}{\mcitedefaultseppunct}\relax
\EndOfBibitem
\bibitem[Donaldson \latin{et~al.}(2021)Donaldson, Schall, and
  Rossi]{Donaldson:2021jz}
Donaldson,~J.~G.; Schall,~P.; Rossi,~L. {Magnetic Coupling in Colloidal
  Clusters for Hierarchical Self-Assembly}. \emph{ACS Nano} \textbf{2021},
  \emph{15}, 4989--4999\relax
\mciteBstWouldAddEndPuncttrue
\mciteSetBstMidEndSepPunct{\mcitedefaultmidpunct}
{\mcitedefaultendpunct}{\mcitedefaultseppunct}\relax
\EndOfBibitem
\bibitem[Rossi \latin{et~al.}(2015)Rossi, Soni, Ashton, Pine, Philipse,
  Chaikin, Dijkstra, Sacanna, and Irvine]{Rossi:2015jf}
Rossi,~L.; Soni,~V.; Ashton,~D.~J.; Pine,~D.~J.; Philipse,~A.~P.;
  Chaikin,~P.~M.; Dijkstra,~M.; Sacanna,~S.; Irvine,~W. T.~M. {Shape-sensitive
  crystallization in colloidal superball fluids}. \emph{Proceedings Of The
  National Academy Of Sciences Of The United States Of America} \textbf{2015},
  \emph{112}, 5286--5290\relax
\mciteBstWouldAddEndPuncttrue
\mciteSetBstMidEndSepPunct{\mcitedefaultmidpunct}
{\mcitedefaultendpunct}{\mcitedefaultseppunct}\relax
\EndOfBibitem
\bibitem[van Anders \latin{et~al.}(2014)van Anders, Klotsa, Ahmed, Engel, and
  Glotzer]{entint}
van Anders,~G.; Klotsa,~D.; Ahmed,~N.~K.; Engel,~M.; Glotzer,~S.~C.
  Understanding shape entropy through local dense packing. \emph{Proc. Natl.
  Acad. Sci. U.S.A.} \textbf{2014}, \emph{111}, E4812--E4821\relax
\mciteBstWouldAddEndPuncttrue
\mciteSetBstMidEndSepPunct{\mcitedefaultmidpunct}
{\mcitedefaultendpunct}{\mcitedefaultseppunct}\relax
\EndOfBibitem
\bibitem[Teich \latin{et~al.}(2016)Teich, van Anders, Klotsa, Dshemuchadse, and
  Glotzer]{Teich:2016ij}
Teich,~E.~G.; van Anders,~G.; Klotsa,~D.; Dshemuchadse,~J.; Glotzer,~S.~C.
  {Clusters of polyhedra in spherical confinement}. \emph{Proceedings Of The
  National Academy Of Sciences Of The United States Of America} \textbf{2016},
  \emph{113}, E669--E678\relax
\mciteBstWouldAddEndPuncttrue
\mciteSetBstMidEndSepPunct{\mcitedefaultmidpunct}
{\mcitedefaultendpunct}{\mcitedefaultseppunct}\relax
\EndOfBibitem
\bibitem[Ellis(2001)]{Ellis:2001bu}
Ellis,~R.~J. {Macromolecular crowding: obvious but underappreciated}.
  \emph{Trends in Biochemical Sciences} \textbf{2001}, \emph{26},
  597--604\relax
\mciteBstWouldAddEndPuncttrue
\mciteSetBstMidEndSepPunct{\mcitedefaultmidpunct}
{\mcitedefaultendpunct}{\mcitedefaultseppunct}\relax
\EndOfBibitem
\bibitem[Hayashi and Carthew(2004)Hayashi, and Carthew]{Hayashi:2004ita}
Hayashi,~T.; Carthew,~R.~W. {Surface mechanics mediate pattern formation in the
  developing retina}. \emph{Nature} \textbf{2004}, \emph{431}, 647--652\relax
\mciteBstWouldAddEndPuncttrue
\mciteSetBstMidEndSepPunct{\mcitedefaultmidpunct}
{\mcitedefaultendpunct}{\mcitedefaultseppunct}\relax
\EndOfBibitem
\bibitem[{\AA}str{\"o}m and Karttunen(2006){\AA}str{\"o}m, and
  Karttunen]{Astrom:2006kq}
{\AA}str{\"o}m,~J.~A.; Karttunen,~M. {Cell aggregation: Packing soft grains}.
  \emph{Physical Review E} \textbf{2006}, \emph{73}, 062301--4\relax
\mciteBstWouldAddEndPuncttrue
\mciteSetBstMidEndSepPunct{\mcitedefaultmidpunct}
{\mcitedefaultendpunct}{\mcitedefaultseppunct}\relax
\EndOfBibitem
\bibitem[Gerba and Betancourt(2017)Gerba, and Betancourt]{Gerba:2017bc}
Gerba,~C.~P.; Betancourt,~W.~Q. {Viral Aggregation: Impact on Virus Behavior in
  the Environment}. \emph{Environmental Science {\&} Technology} \textbf{2017},
  \emph{51}, 7318--7325\relax
\mciteBstWouldAddEndPuncttrue
\mciteSetBstMidEndSepPunct{\mcitedefaultmidpunct}
{\mcitedefaultendpunct}{\mcitedefaultseppunct}\relax
\EndOfBibitem
\bibitem[Cines \latin{et~al.}(2014)Cines, Lebedeva, Nagaswami, Hayes,
  Massefski, Litvinov, Rauova, Lowery, and Weisel]{Cines:2014fka}
Cines,~D.~B.; Lebedeva,~T.; Nagaswami,~C.; Hayes,~V.; Massefski,~W.;
  Litvinov,~R.~I.; Rauova,~L.; Lowery,~T.~J.; Weisel,~J.~W. {Clot contraction:
  compression of erythrocytes into tightly packed polyhedra and redistribution
  of platelets and fibrin}. \emph{Blood} \textbf{2014}, \emph{123},
  1596--1603\relax
\mciteBstWouldAddEndPuncttrue
\mciteSetBstMidEndSepPunct{\mcitedefaultmidpunct}
{\mcitedefaultendpunct}{\mcitedefaultseppunct}\relax
\EndOfBibitem
\bibitem[Jiao \latin{et~al.}(2009)Jiao, Stillinger, and Torquato]{Jiao:2009jw}
Jiao,~Y.; Stillinger,~F.~H.; Torquato,~S. {Optimal packings of superballs}.
  \emph{Physical Review E} \textbf{2009}, \emph{79}, 041309\relax
\mciteBstWouldAddEndPuncttrue
\mciteSetBstMidEndSepPunct{\mcitedefaultmidpunct}
{\mcitedefaultendpunct}{\mcitedefaultseppunct}\relax
\EndOfBibitem
\bibitem[Jiao \latin{et~al.}(2011)Jiao, Stillinger, and Torquato]{Jiao:2011jh}
Jiao,~Y.; Stillinger,~F.~H.; Torquato,~S. {Erratum: Optimal packings of
  superballs}. \emph{Physical Review E} \textbf{2011}, \emph{84}, 069902\relax
\mciteBstWouldAddEndPuncttrue
\mciteSetBstMidEndSepPunct{\mcitedefaultmidpunct}
{\mcitedefaultendpunct}{\mcitedefaultseppunct}\relax
\EndOfBibitem
\bibitem[Ni \latin{et~al.}(2012)Ni, Gantapara, de~Graaf, van Roij, and
  Dijkstra]{Ni:2012bc}
Ni,~R.; Gantapara,~A.~P.; de~Graaf,~J.; van Roij,~R.; Dijkstra,~M. {Phase
  diagram of colloidal hard superballs: from cubes via spheres to octahedra}.
  \emph{Soft Matter} \textbf{2012}, \emph{8}, 8826\relax
\mciteBstWouldAddEndPuncttrue
\mciteSetBstMidEndSepPunct{\mcitedefaultmidpunct}
{\mcitedefaultendpunct}{\mcitedefaultseppunct}\relax
\EndOfBibitem
\bibitem[Meijer \latin{et~al.}(2017)Meijer, Pal, Ouhajji, Lekkerkerker,
  Philipse, and Petukhov]{Meijer:2017hr}
Meijer,~J.-M.; Pal,~A.; Ouhajji,~S.; Lekkerkerker,~H. N.~W.; Philipse,~A.~P.;
  Petukhov,~A.~V. {Observation of solid-solid transitions in 3D crystals of
  colloidal superballs}. \emph{Nature Communications} \textbf{2017}, \emph{8},
  14352\relax
\mciteBstWouldAddEndPuncttrue
\mciteSetBstMidEndSepPunct{\mcitedefaultmidpunct}
{\mcitedefaultendpunct}{\mcitedefaultseppunct}\relax
\EndOfBibitem
\bibitem[Stober \latin{et~al.}(1968)Stober, Fink, and Bohn]{Stober:1968tt}
Stober,~W.; Fink,~A.; Bohn,~E. {Controlled Growth of Monodisperse Silica
  Spheres in the Micron Size Range}. \emph{Journal of Colloid and Interface
  Science} \textbf{1968}, \emph{26}, 62--69\relax
\mciteBstWouldAddEndPuncttrue
\mciteSetBstMidEndSepPunct{\mcitedefaultmidpunct}
{\mcitedefaultendpunct}{\mcitedefaultseppunct}\relax
\EndOfBibitem
\bibitem[Cho \latin{et~al.}(2005)Cho, Yi, Kim, Pine, and Yang]{Cho:2005cp}
Cho,~Y.-S.; Yi,~G.-R.; Kim,~S.-H.; Pine,~D.~J.; Yang,~S.-M. {Colloidal Clusters
  of Microspheres from Water-in-Oil Emulsions}. \emph{Chemistry Of Materials}
  \textbf{2005}, \emph{17}, 5006--5013\relax
\mciteBstWouldAddEndPuncttrue
\mciteSetBstMidEndSepPunct{\mcitedefaultmidpunct}
{\mcitedefaultendpunct}{\mcitedefaultseppunct}\relax
\EndOfBibitem
\bibitem[Soligno \latin{et~al.}(2018)Soligno, Dijkstra, and van
  Roij]{Soligno:2018dt}
Soligno,~G.; Dijkstra,~M.; van Roij,~R. {Self-assembly of cubic colloidal
  particles at fluid{\textendash}fluid interfaces by hexapolar capillary
  interactions}. \emph{Soft Matter} \textbf{2018}, \emph{14}, 42--60\relax
\mciteBstWouldAddEndPuncttrue
\mciteSetBstMidEndSepPunct{\mcitedefaultmidpunct}
{\mcitedefaultendpunct}{\mcitedefaultseppunct}\relax
\EndOfBibitem
\bibitem[Soligno \latin{et~al.}(2016)Soligno, Dijkstra, and van
  Roij]{Soligno:2016hs}
Soligno,~G.; Dijkstra,~M.; van Roij,~R. {Self-Assembly of Cubes into 2D
  Hexagonal and Honeycomb Lattices by Hexapolar Capillary Interactions}.
  \emph{Physical Review Letters} \textbf{2016}, \emph{116}, 258001--6\relax
\mciteBstWouldAddEndPuncttrue
\mciteSetBstMidEndSepPunct{\mcitedefaultmidpunct}
{\mcitedefaultendpunct}{\mcitedefaultseppunct}\relax
\EndOfBibitem
\bibitem[Manoharan(2003)]{Manoharan:2003hb}
Manoharan,~V.~N. {Dense Packing and Symmetry in Small Clusters of
  Microspheres}. \emph{Science} \textbf{2003}, \emph{301}, 483--487\relax
\mciteBstWouldAddEndPuncttrue
\mciteSetBstMidEndSepPunct{\mcitedefaultmidpunct}
{\mcitedefaultendpunct}{\mcitedefaultseppunct}\relax
\EndOfBibitem
\bibitem[Manoharan and Pine(2004)Manoharan, and Pine]{Manoharan:2004vs}
Manoharan,~V.; Pine,~D. {Building materials by packing spheres}. \emph{MRS
  Bulletin} \textbf{2004}, \emph{29}, 91--95\relax
\mciteBstWouldAddEndPuncttrue
\mciteSetBstMidEndSepPunct{\mcitedefaultmidpunct}
{\mcitedefaultendpunct}{\mcitedefaultseppunct}\relax
\EndOfBibitem
\bibitem[Lauga and Brenner(2004)Lauga, and Brenner]{Lauga:2004im}
Lauga,~E.; Brenner,~M.~P. {Evaporation-Driven Assembly of Colloidal Particles}.
  \emph{Physical Review Letters} \textbf{2004}, \emph{93}, 237--4\relax
\mciteBstWouldAddEndPuncttrue
\mciteSetBstMidEndSepPunct{\mcitedefaultmidpunct}
{\mcitedefaultendpunct}{\mcitedefaultseppunct}\relax
\EndOfBibitem
\bibitem[Rossi \latin{et~al.}(2018)Rossi, Donaldson, Meijer, Petukhov,
  Kleckner, Kantorovich, Irvine, Philipse, and Sacanna]{Rossi:2018ef}
Rossi,~L.; Donaldson,~J.~G.; Meijer,~J.-M.; Petukhov,~A.~V.; Kleckner,~D.;
  Kantorovich,~S.~S.; Irvine,~W. T.~M.; Philipse,~A.~P.; Sacanna,~S.
  {Self-organization in dipolar cube fluids constrained by competing
  anisotropies}. \emph{Soft Matter} \textbf{2018}, \emph{14}, 1080--1087\relax
\mciteBstWouldAddEndPuncttrue
\mciteSetBstMidEndSepPunct{\mcitedefaultmidpunct}
{\mcitedefaultendpunct}{\mcitedefaultseppunct}\relax
\EndOfBibitem
\bibitem[Anderson \latin{et~al.}(2015)Anderson, {Eric Irrgang}, and
  Glotzer]{Anderson2015a}
Anderson,~J.~A.; {Eric Irrgang},~M.; Glotzer,~S.~C. {Scalable Metropolis Monte
  Carlo for simulation of hard shapes}. \emph{Computer Physics Communications}
  \textbf{2015}, \emph{204}, 21--30\relax
\mciteBstWouldAddEndPuncttrue
\mciteSetBstMidEndSepPunct{\mcitedefaultmidpunct}
{\mcitedefaultendpunct}{\mcitedefaultseppunct}\relax
\EndOfBibitem
\bibitem[Anderson \latin{et~al.}(2008)Anderson, Lorenz, and
  Travesset]{Anderson2008b}
Anderson,~J.~A.; Lorenz,~C.~D.; Travesset,~A. {General purpose molecular
  dynamics simulations fully implemented on graphics processing units}.
  \emph{Journal of Computational Physics} \textbf{2008}, \emph{227},
  5342--5359\relax
\mciteBstWouldAddEndPuncttrue
\mciteSetBstMidEndSepPunct{\mcitedefaultmidpunct}
{\mcitedefaultendpunct}{\mcitedefaultseppunct}\relax
\EndOfBibitem
\bibitem[Damasceno \latin{et~al.}(2012)Damasceno, Engel, and Glotzer]{trunctet}
Damasceno,~P.~F.; Engel,~M.; Glotzer,~S.~C. {Crystalline Assemblies and Densest
  Packings of a Family of Truncated Tetrahedra and the Role of Directional
  Entropic Forces}. \emph{ACS Nano} \textbf{2012}, \emph{6}, 609--614\relax
\mciteBstWouldAddEndPuncttrue
\mciteSetBstMidEndSepPunct{\mcitedefaultmidpunct}
{\mcitedefaultendpunct}{\mcitedefaultseppunct}\relax
\EndOfBibitem
\bibitem[Damasceno \latin{et~al.}(2012)Damasceno, Engel, and Glotzer]{zoopaper}
Damasceno,~P.~F.; Engel,~M.; Glotzer,~S.~C. {Predictive Self-Assembly of
  Polyhedra into Complex Structures}. \emph{Science} \textbf{2012}, \emph{337},
  453--457\relax
\mciteBstWouldAddEndPuncttrue
\mciteSetBstMidEndSepPunct{\mcitedefaultmidpunct}
{\mcitedefaultendpunct}{\mcitedefaultseppunct}\relax
\EndOfBibitem
\bibitem[de~Nijs \latin{et~al.}(2014)de~Nijs, Dussi, Smallenburg, Meeldijk,
  Groenendijk, Filion, Imhof, van Blaaderen, and Dijkstra]{deNijs:2014fb}
de~Nijs,~B.; Dussi,~S.; Smallenburg,~F.; Meeldijk,~J.~D.; Groenendijk,~D.~J.;
  Filion,~L.; Imhof,~A.; van Blaaderen,~A.; Dijkstra,~M. {Entropy-driven
  formation of large icosahedral colloidal clusters by spherical confinement}.
  \emph{Nature Materials} \textbf{2014}, \emph{14}, 56--60\relax
\mciteBstWouldAddEndPuncttrue
\mciteSetBstMidEndSepPunct{\mcitedefaultmidpunct}
{\mcitedefaultendpunct}{\mcitedefaultseppunct}\relax
\EndOfBibitem
\bibitem[Glaser \latin{et~al.}(2015)Glaser, Nguyen, Anderson, Lui, Spiga,
  Millan, Morse, and Glotzer]{Glaser2015}
Glaser,~J.; Nguyen,~T.~D.; Anderson,~J.~A.; Lui,~P.; Spiga,~F.; Millan,~J.~A.;
  Morse,~D.~C.; Glotzer,~S.~C. {Strong scaling of general-purpose molecular
  dynamics simulations on GPUs}. \emph{Computer Physics Communications}
  \textbf{2015}, \emph{192}, 97--107\relax
\mciteBstWouldAddEndPuncttrue
\mciteSetBstMidEndSepPunct{\mcitedefaultmidpunct}
{\mcitedefaultendpunct}{\mcitedefaultseppunct}\relax
\EndOfBibitem
\bibitem[Adorf \latin{et~al.}(2018)Adorf, Dodd, Ramasubramani, and
  Glotzer]{Adorf2016a}
Adorf,~C.~S.; Dodd,~P.~M.; Ramasubramani,~V.; Glotzer,~S.~C. {Simple data and
  workflow management with the signac framework}. \emph{Computational Materials
  Science} \textbf{2018}, \emph{146}, 220--229\relax
\mciteBstWouldAddEndPuncttrue
\mciteSetBstMidEndSepPunct{\mcitedefaultmidpunct}
{\mcitedefaultendpunct}{\mcitedefaultseppunct}\relax
\EndOfBibitem
\bibitem[Adorf \latin{et~al.}(2017)Adorf, Dodd, Ramasubramani, Swerdlow,
  Glaser, and Dice]{Adorf2017a}
Adorf,~C.~S.; Dodd,~P.~M.; Ramasubramani,~V.; Swerdlow,~B.; Glaser,~J.;
  Dice,~B. csadorf/signac v0.9.2. \textbf{2017}, \relax
\mciteBstWouldAddEndPunctfalse
\mciteSetBstMidEndSepPunct{\mcitedefaultmidpunct}
{}{\mcitedefaultseppunct}\relax
\EndOfBibitem
\bibitem[Engel(2021)]{injavis}
Engel,~M. INJAVIS — INteractive JAva VISualization. 2021\relax
\mciteBstWouldAddEndPuncttrue
\mciteSetBstMidEndSepPunct{\mcitedefaultmidpunct}
{\mcitedefaultendpunct}{\mcitedefaultseppunct}\relax
\EndOfBibitem
\bibitem[Marechal \latin{et~al.}(2012)Marechal, Zimmermann, and
  L{\"o}wen]{Marechal2012}
Marechal,~M.; Zimmermann,~U.; L{\"o}wen,~H. {Freezing of parallel hard cubes
  with rounded edges}. \emph{Journal of Chemical Physics} \textbf{2012},
  \emph{136}, 144506--1--14\relax
\mciteBstWouldAddEndPuncttrue
\mciteSetBstMidEndSepPunct{\mcitedefaultmidpunct}
{\mcitedefaultendpunct}{\mcitedefaultseppunct}\relax
\EndOfBibitem
\end{mcitethebibliography}
\end{document}

\title{Supporting Information:
	\\Shape and Interaction Decoupling for Colloidal Pre-Assembly}
\author[1,2]{L. Baldauf}
\author[3,4]{E. G. Teich}
\author[1]{P. Schall}
\author[5,6]{G. van Anders $^\star$}
\author[7]{Laura Rossi $^\star$}

\affil[1]{Institute of Physics, University of Amsterdam, 1098XH Amsterdam, The Netherlands.}
\affil[2]{Current Address: Department of Bionanoscience, Delft University ofTechnology, 2629 HZ Delft, The Netherlands.}
\affil[3]{Applied Physics Program, University of Michigan, Ann Arbor, MI 48109,USA.}
\affil[4]{Current Address: Department of Bioengineering, University ofPennsylvania, Philadelphia, PA 19104, USA.}
\affil[5]{Physics, University of Michigan, Ann Arbor, MI 48109, USA.}
\affil[6]{Physics, Engineering Physics, and Astronomy, Queen’s University, Kingston ON K7L3N6, Canada.}
\affil[7]{Department of Chemical Engineering, Delft University of Technology, 2629 HZ Delft, The Netherlands.}

\renewcommand\Authands{ and }

\maketitle

\begin{footnotesize}
	$^\star$ gva@queensu.ca, l.rossi@tudelft.nl
	\\
\end{footnotesize}

\doublespacing

\newpage

	%------------------
% MAIN TEXT
%------------------
\section*{Supporting data}

\begin{figure}[h]	\renewcommand{\thefigure}{S\arabic{figure}}
	\centering
	\includegraphics[scale=0.8]{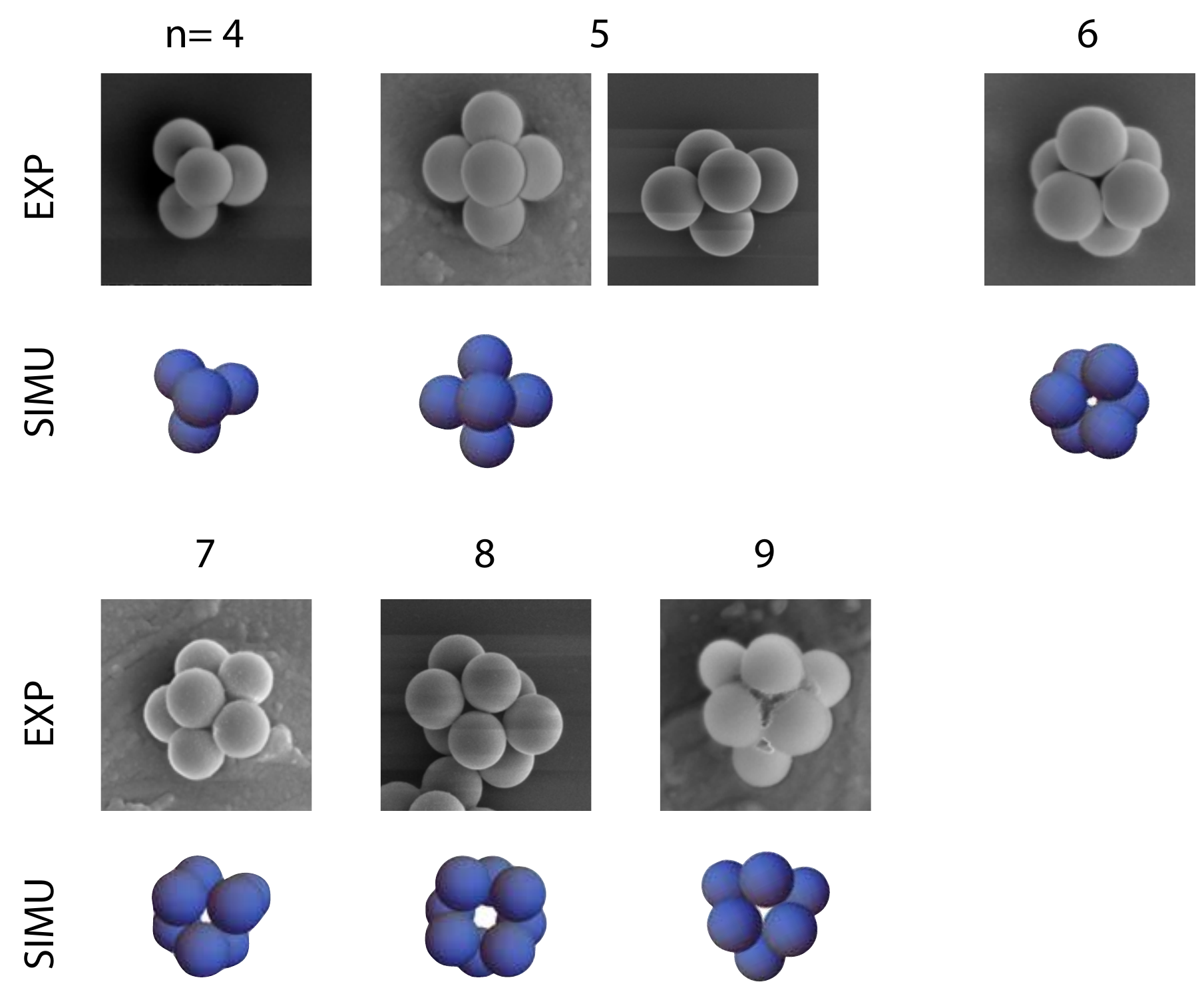}
	\caption{SEM images of clusters from water-in-oil emulsions obtained using spherical particles ($d=648$ nm  and $1.2$ \si{\micro\meter}) and computer simulations of clusters with the same number of constituent spherical particles.}  
    \label{sphclus}
\end{figure}

%===============
% Hierarchical self-assembly
%===============
\noindent \textbf{Sphere Clusters}
\\
\\
To test the validity of the experimental procedure we reproduced clusters of spherical particles using two different samples of silica spheres: one with diameter $d=466$ \si{\nano\meter} synthesized by us and one with diameter $d=1.2$ \si{\micro\meter} purchased from Bangs Laboratories Inc. The collective results are reported in Figure~\ref{sphclus}. The results obtained show that we have successfully reproduced the procedure reported by Cho \textit{et al.} \cite{Cho:2005cp}. In our experiments for clusters with $N=5$ components we find, however, both a triangular dypiramid, which structure minimizes the second-moment of the mass distribution, and a square pyramid which has not been reported so far in emulsion experiments. 

The experimental observations match very well with the clusters obtained by computer simulations, with the exception of $N=8$, where simulations show a twisted square configuration, but experiments find a snub disphenoid, and $N=7$, where no match was found between our experiments and simulations. In our experiments we observe only one of the isomers (pentagonal dipyramid) reported for clusters generated from w-i-o emulsions \cite{Cho:2005cp}, whereas the other isomer (tetramer-on-trimer) is predicted by our model.

In the case of $N=5$, 30 of our 50 replica simulations resulted in the square pyramid, while the remaining 20 resulted in configurations resembling triangular bipyramids. We found the densest cluster to be a square pyramid, but density distributions for the square pyramid and triangular bipyramid structures were approximately equal, and the density of the densest square pyramid structure was only larger than the density of the densest triangular bipyramid structure by $\approx 1.8 \times 10^{-6}$. It is known that the $N=5$ spherical code, the densest configuration of equal radius circles on the surface of a sphere, has a continuum of solutions between the square pyramid and the triangular bipyramid \cite{Phillips2012}. Authors of Ref. \citenum{Phillips2012} found that, for 5 spheres packing around a central smaller sphere, the square pyramid is entropically favorable over the triangular bipyramid, as it results in the highest vibrational freedom for the particles. Thus, the degeneracy that we see in these structures, both in experiments and simulations, makes sense, as does the slight preference for the square pyramid shown by our simulations.

%------------------
% FIGURE S1
%------------------

\newpage

\section*{Supporting Figures}

%------------------
% FIGURE S2
%------------------
\begin{figure}[h!]\renewcommand{\thefigure}{S\arabic{figure}}
	\centering
	\includegraphics[scale=0.7]{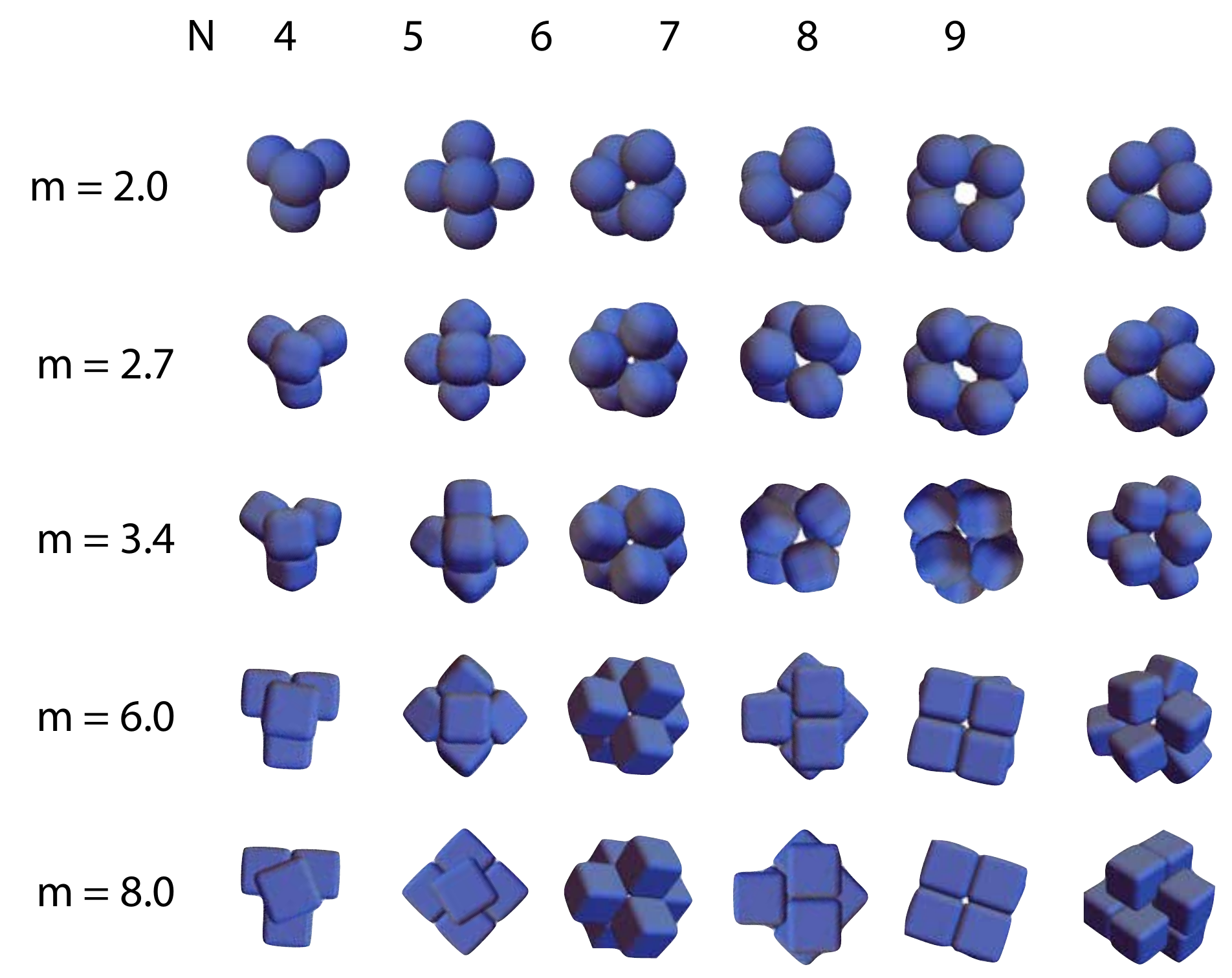}
	\caption{Clusters generated by computer simulation for superballs with various $m$ values.}  
    \label{simuhighm}
\end{figure}

\hspace{3cm}

\providecommand{\latin}[1]{#1}
\makeatletter
\providecommand{\doi}
  {\begingroup\let\do\@makeother\dospecials
  \catcode`\{=1 \catcode`\}=2 \doi@aux}
\providecommand{\doi@aux}[1]{\endgroup\texttt{#1}}
\makeatother
\providecommand*\mcitethebibliography{\thebibliography}
\csname @ifundefined\endcsname{endmcitethebibliography}
  {\let\endmcitethebibliography\endthebibliography}{}